\documentclass{elsart}
\usepackage[normalem]{ulem}
\usepackage{graphicx}
\usepackage{latexsym,amssymb,amsmath,amssymb,xspace,theorem}
\usepackage{multicol}
\usepackage{t1enc}
\usepackage{epic,eepic}
\usepackage{algorithmic}
\usepackage{subfigure}
\theoremstyle{plain} \theoremheaderfont{\scshape}

\newtheorem{Thm}{\bf Theorem}
\newtheorem{Lem}[Thm]{\bf Lemma}
\newtheorem{Clm}{Claim}[Thm]

\newtheorem{Conj}[Thm]{{\bf Conjecture}}

\newtheorem{Cor}[Thm]{ \bf Corollary}
{\theorembodyfont{\rmfamily}

}

\newenvironment{Prf}{{\bf \noindent Proof } }{\hfill$\square$\\}
\newenvironment{PrfClaim}{{\bf Proof }}{{\hfill\tiny{$\blacksquare$\\}}}

\newcommand{\ignore}[1]{}

\newcommand{\cqfd}{\unskip\kern 6pt\penalty 500
\raise -2pt\hbox{\vrule\vbox to 10pt{\hrule width 4pt
\vfill\hrule}\vrule}\par}

\input{epsf.sty}
\begin{document}
\begin{frontmatter}


\title{On Fan Raspaud conjecture}
\author{J.L. Fouquet and J.M. Vanherpe}

\address{L.I.F.O., Facult\'e des Sciences, B.P. 6759 \\
Universit\'e d'Orl\'eans, 45067 Orl\'eans Cedex 2, FR}
\begin{abstract}
A conjecture of Fan and Raspaud \cite{FanRas} asserts that every
bridgeless cubic graph contains three perfect matchings with empty
intersection. Kaiser and Raspaud \cite{KaiRas} suggested a possible
approach to this problem based on the concept of a balanced join in
an embedded graph. We give here some new results concerning this
conjecture and prove that a minimum counterexample must have at
least $32$ vertices.
\end{abstract}
\end{frontmatter}

\section{Introduction}
Fan and Raspaud \cite{FanRas} conjectured that any bridgeless cubic
graph can be provided with three perfect matchings with empty
intersection (we shall say also {\em non intersecting perfect
matchings}).
\begin{Conj}\cite{FanRas} \label{Conjecture:FanRaspaud} Every
bridgeless cubic graph contains perfect matching $M_1$, $M_2$, $M_3$
such that
$$M_1 \cap M_2 \cap M_3 = \emptyset$$
\end{Conj}

This conjecture seems to be originated independently by Jackson.
Goddyn \cite{God} indeed mentioned this problem proposed by Jackson
for $r-$graphs ($r-$regular graphs with an even number of vertices
such that all odd cuts have size at least $r$, as defined by Seymour
\cite{Sey}) in the proceedings of a joint summer research conference
on graphs minors which dates back 1991.

\begin{Conj}\cite{God} \label{Conjecture:Jackson}There exists $k \geq 2$ such
that any r-graph contains $k+1$ perfect matchings with empty
intersection.
\end{Conj}
Seymour \cite{Sey} conjectured that:
\begin{Conj}\cite{Sey} \label{Conjecture:Seymour}If $r \geq 4$ then any
r-graph has a perfect matching whose deletion yields an (r-1)-graph.
\end{Conj}

Hence Seymour's conjecture leads to a  specialized form of Jackson's
conjecture when dealing with cubic bridgeless graphs and the Fan
Raspaud conjecture appears as a refinement of Jackson's conjecture.

 A {\em join} in a graph $G$ is a set $J \subseteq
E(G)$ such that the degree of every vertex in G has the same parity
as its degree in the graph $(V (G), J)$. A perfect matching being a
particular join in a cubic graph Kaiser and Raspaud conjectured in
\cite{KaiRas}

\begin{Conj}\cite{KaiRas} \label{Conjecture:KaiserRaspaud} Every
bridgeless cubic graph admits two perfect matching $M_1$, $M_2$ and
a join $J$ such that
$$M_1 \cap M_2 \cap J = \emptyset$$
\end{Conj}

The {\em oddness} of a cubic graph $G$ is the minimum number of odd
circuits in a 2-factor of $G$. Conjecture
\ref{Conjecture:FanRaspaud} being obviously true for cubic graphs
with chromatic index $3$, we shall be concerned here by bridgeless
cubic graphs with chromatic index $4$. Hence any 2-factor of such a
graph has at least two odd cycles. The class of  bridgeless cubic
graphs with oddness two is, in some sense, the "easiest" class to
manage with in order to tackle some well known conjecture.  In
\cite{KaiRas} Kaiser and Raspaud proved that Conjecture
\ref{Conjecture:KaiserRaspaud} holds true for bridgeless cubic graph
of oddness  two. Their proof is based on the notion of {\em balanced
join} in the multigraph obtained in contracting the cycles of a two
factor. Using an equivalent formulation of this notion in the next
section, we shall see that we can get some new results on Conjecture
\ref{Conjecture:FanRaspaud} with the help of this technique.

For basic graph-theoretic terms, we refer the reader to Bondy and
Murty \cite{BonMur}.

\section{Preliminary results}
Let $M$ be a perfect matching of a  cubic graph and let $\mathcal
C=\{C_1,C_2 \ldots C_k\}$ be the 2-factor $G-M$. $A \subseteq M$ is
a {\em balanced} $M-$matching  whenever there is a perfect matching
$M'$ such that $M \cap M' =A$. That means that each odd cycle of
$\mathcal C$ is incident to at least one edge in $A$ and the
subpaths determined by the ends of $M'$ on the cycles of $\mathcal
C$ incident to $A$ have odd lengths.

In the following example, $M$ is the perfect matching (thick edges)
of the Petersen graph. Taking any edge ($ab$ by example) of this
perfect matching we are led to a balanced $M-$matching since the two
cycles of length $5$ give rise to two paths of length $5$ (we have
"opened" these paths closed to $a$ and $b$). Remark that given a
perfect matching $M$ of a bridgeless cubic graph, $M$ is obviously a
balanced $M-$matching.

\begin{figure}[t]
\centering \epsfsize=0.35 \hsize \noindent \epsfbox{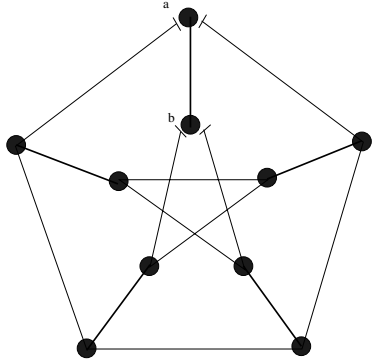}
\caption{A balanced $M-$matching} \label{Figure:Petersen}
\end{figure}

Kaiser and Raspaud \cite{KaiRas} introduced this notion via the
notion of {\em balanced join} in  the context of a combinatorial
representation of graphs embedded on surfaces. They remarked that a
natural approach to the Fan Raspaud conjecture would require finding
two disjoint balanced joins and hence two balanced $M-$matchings for
some perfect matching $M$. In fact  Conjecture
\ref{Conjecture:FanRaspaud} and balanced matching are related by the
following lemma

\begin{Lem}\label{Lemma:FondamentalDisjointsBalanced}
A bridgeless cubic graph contains $3$ non intersecting perfect
matching if and only if there is a perfect matching $M$ and two
balanced disjoint balanced $M-$matchings.
\end{Lem}
\begin{Prf} Assume that $M_1$, $M_2$, $M_3$ are three perfect
matchings of $G$ such that $M_1 \cap M_2 \cap M_3 = \emptyset$. Let
$M=M_1$, $A=M_1 \cap M_2$ and $B=M_1 \cap M_3$. Since $A\cap B=M_1
\cap M_2 \cap M_3$, $A$ and $B$ are two balanced $M-$matchings with
empty intersection.

Conversely, assume that $M$ is a perfect matching and that $A$ and
$B$ are two balanced $M-$matchings with empty intersection. Let
$M_1=M$, $M_2$ be a perfect matching such that $M_2 \cap M_1=A$
 and $M_3$ be a perfect
matching such that $M_3 \cap M_1=B$. We have $M_1 \cap M_2 \cap
M_3=A \cap B$ and the three perfect matchings $M_1$, $M_2$ and $M_3$
have an empty intersection.
\end{Prf}

 The following theorem is a corollary of Edmond's Matching Polyhedron
Theorem \cite{Edm}. A simple proof is given by Seymour in
\cite{Sey}.

\begin{Thm} \label{Theorem:EdmondsSeymour} Let $G$ be an $r$-graph. Then there is an integer $p$
and a family $\mathcal M$ of perfect matchings such that each edge
of $G$ is contained in precisely $p$ members of $\mathcal M$.
\end{Thm}

\begin{Lem}\label{Lemma:TwoEdgesAvoiding} Let $G$ be a bridgeless
cubic graph and let $e=uv$ and $e'=u'v'$ be two edges of $G$. Then
there exists a perfect matching avoiding these two edges.
\end{Lem}
\begin{Prf}  Remark that a bridgeless cubic graph is a $3$-graph as defined by Seymour.
Applying Theorem \ref{Theorem:EdmondsSeymour}, let $\mathcal M$ be a
set of perfect matching such that each edge of $G$ is contained in
precisely $p$ members of $\mathcal M$ (for some fixed integer $p
\geq 1$).

Assume first that $e$ and $e'$ have a common end vertex (say $u$).
Then $ u$ is incident to a third edge $e"$. Any perfect matching
using $e"$ avoids $e$ and $e'$.

When $e$ and $e'$ have no common end then, let $f$ and $g$ be the
two edges incident with $u$. Assume that any perfect matching using
$f$ or $g$ contains also the edge $e'$. Then $e'$ is contained in
$2p$ members of $\mathcal M$, impossible. Hence some perfect
matchings using $f$ or $g$ must avoid $e'$, as claimed.
\end{Prf}

It can be pointed out that Lemma \ref{Lemma:TwoEdgesAvoiding} is not
extendable, so easily, to a larger set of edges. Indeed, a corollary
of Theorem \ref{Theorem:EdmondsSeymour} asserts that $\mathcal M$
(the family of perfect matching considered) intersects each $3-$edge
cut in exactly one edge. Hence for such a $3-$edge cut, there is no
perfect matching in $\mathcal M$ avoiding this set.

Let $C$ be an odd cycle and let $T=\{x,y,z\}$ a set of three
distinct vertices of $C$. We shall say that $C$ is a {\em balanced
triple} when the three subpaths of $C$ determined by $T$ have odd
lengths.

\begin{figure}[t]
\centering \epsfsize=0.25 \hsize \noindent
\epsfbox{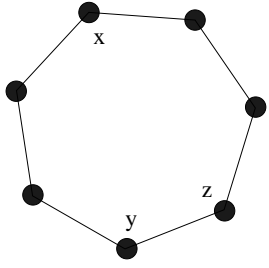} \caption{A balanced triple}
\label{Figure:BalancedTriple}
\end{figure}

Let $C=x_0x_1 \ldots x_{2k}$ be an odd cyle of length at least $7$.
Assume that its vertex set is coloured with three colours $1$, $2$
and $3$ such that $2 \leq |A_1| \leq |A_2| \leq |A_3|$, $A_i$
denoting the set of vertices coloured with $i$, $i =1,2,3$. Then we
shall say that $C$ is {\em good odd cycle}.

\begin{Lem}\label{Lemma:DisjointBalancedTriples}  Any good odd cycle $C$
contains two disjoint balanced triples $T$ and $T'$ intersecting
each colour exactly once.
\end{Lem}
\begin{Prf} We
shall prove this lemma by induction on $|C|$.

Assume first that $C$ has length $7$. Then $A_1$ and $A_2$ have
exactly two vertices while $A_3$ must have $3$ vertices. We can
distinguish, up to isomorphism, $9$ subcases

\begin{enumerate}
  \item $A_3=\{x_0,x_1,x_2\}$ $A_1=\{x_3,x_4\}$ and
  $A_2=\{x_5,x_6\}$ then $T=\{x_0,x_3,x_6\}$ and $T'=\{x_1,x_4,x_5\}$ are  two
  disjoint balanced triples.
  \item $A_3=\{x_0,x_1,x_2\}$ $A_1=\{x_3,x_5\}$ and
  $A_2=\{x_4,x_6\}$ then $T=\{x_2,x_3,x_4\}$ and $T'=\{x_5,x_6,x_0\}$ are  two
  disjoint balanced triples.
  \item$A_3=\{x_0,x_1,x_2\}$ $A_1=\{x_3,x_6\}$ and
  $A_2=\{x_4,x_5\}$ then $T=\{x_2,x_3,x_4\}$ and $T'=\{x_5,x_6,x_0\}$ are  two
  disjoint balanced triples.
  \item$A_3=\{x_0,x_1,x_3\}$ $A_1=\{x_2,x_4\}$ and
  $A_2=\{x_5,x_6\}$ then $T=\{x_1,x_4,x_5\}$ and $T'=\{x_2,x_3,x_6\}$ are  two
  disjoint balanced triples.
  \item$A_3=\{x_0,x_1,x_3\}$ $A_1=\{x_2,x_5\}$ and
  $A_2=\{x_4,x_6\}$ then $T=\{x_1,x_4,x_5\}$ and $T'=\{x_2,x_3,x_6\}$ are  two
  disjoint balanced triples.
  \item$A_3=\{x_0,x_1,x_3\}$ $A_1=\{x_2,x_6\}$ and
  $A_2=\{x_4,x_5\}$ then $T=\{x_1,x_0,x_6\}$ and $T'=\{x_2,x_3,x_4\}$ are  two
  disjoint balanced triples.
  \item$A_3=\{x_0,x_1,x_4\}$ $A_1=\{x_2,x_3\}$ and
  $A_2=\{x_5,x_6\}$ then $T=\{x_1,x_2,x_5\}$ and $T'=\{x_0,x_3,x_6\}$ are  two
  disjoint balanced triples.
  \item$A_3=\{x_0,x_1,x_4\}$ $A_1=\{x_2,x_5\}$ and
  $A_2=\{x_3,x_6\}$ then $T=\{x_1,x_2,x_3\}$ and $T'=\{x_0,x_5,x_6\}$ are  two
  disjoint balanced triples.
  \item$A_3=\{x_0,x_1,x_4\}$ $A_1=\{x_2,x_6\}$ and
  $A_2=\{x_3,x_5\}$ then $T=\{x_1,x_2,x_3\}$ and $T'=\{x_0,x_5,x_6\}$ are  two
  disjoint balanced triples.
\end{enumerate}

Assume that $C$ is a good odd cycle of length at least $9$ and
assume that the property holds for any good odd cycle of length
$|C|-2$.

\begin{Clm} \label{Claim:Claim1DisjointBalancedTriples}
If $C$ has  two consecutive vertices $x_j$ and $x_{j+1}$ ($j$ being
taken modulo $2k$) in the same set $A_i$ ($i=1,2$ or $3$) such that
$|A_i| \geq 4$, then the property holds.
\end{Clm}
\begin{PrfClaim} Assume that $C$ has two consecutive
vertices $x_j$ and $x_{j+1}$  in the same set $A_i$ ($i=1,2$ or $3$)
such that $|A_i| \geq 4$, then delete $x_j$ and $x_{j+1}$ and add
the edge $x_{j-1}x_{j+2}$. We get hence a good odd cycle $C'$ of
length $|C|-2$. $C'$ has two disjoint balanced triples $T$ and $T'$
by induction hypothesis and we can check that these two triples are
also balanced in $C$ since the edge $x_{j-1}x_{j+2}$ is replaced by
the path $x_{j-1}x_jx_{j+1}x_{j+2}$ in $C$.
\end{PrfClaim}

\begin{Clm} \label{Claim:Claim2DisjointBalancedTriples}
If $C$ has  two consecutive vertices $x_j$ and $x_{j+1}$ ($j$ being
taken modulo $2k$) one of them being in $A_i$ while the other is in
$A_{i'}$ ($i \not = i'  \in \{1,2,3\}$), then the property holds as
soon as $|A_i| \geq 3$ and $|A_{i'}| \geq 3$.
\end{Clm}
\begin{PrfClaim} Use the same trick as in the proof of Claim
\ref{Claim:Claim1DisjointBalancedTriples}
\end{PrfClaim}

If $A_3 \geq 4$, we can suppose, by Claim
\ref{Claim:Claim1DisjointBalancedTriples} that no two vertices of
$A_3$ are consecutive on $C$. When $x \in A_3$, $x'$ (its succesor
in the natural ordering) is in $A_1$ or $A_2$. By Claim
\ref{Claim:Claim2DisjointBalancedTriples}, the vertices in $A_3$
have at most two successors in $A_1$ and at most two successors in
$A_2$. Hence we must have $|A_3| = 4$ and $|A_2|=|A_3|=2$,
impossible. If $|A_3|=3$ then we must have $|A_2|=|A_3|=3$  since
$C$ has length $9$. In that case we certainly have two consecutive
vertices with distinct colours and we can apply the above claim
\ref{Claim:Claim1DisjointBalancedTriples}.
\end{Prf}

Let $C$ be an even cycle and let $P=\{x,y\}$ a set of two distinct
vertices of $C$. We shall say that $C$ is a {\em balanced pair} when
the two subpaths of $C$ determined by $P$ have odd lengths.

Let $C=x_0x_1 \ldots x_{2k-1}$ be an even cyle of length at least
$4$. Assume that its vertex set is coloured with three colours $1$,
$2$ and $3$. Let $A_i$  be  the set of vertices coloured with $i$,
$i =1,2,3$. Assume that  $|A_i|=0$ or $1$ for at most one colour,
then we shall say that $C$ is {\em good even cycle}.

\begin{Lem}\label{Lemma:DisjointBalancedPairs}  Any good even cycle $C$
contains two disjoint balanced pairs $P_i$ and $P'_i$ intersecting
$A_i$  exactly once each as soon as $A_i$ has at least two vertices
($i=1,2,3$).
\end{Lem}
\begin{Prf}
We prove the lemma for $i=1$. Assume that $|A_1| \geq 2$ and $|A_2|
\geq 2$. Assume that $x_0$ is a vertex in $A_2$ and let $x_i$ be the
first vertex in $A_1$, $x_j$ be the last vertex in $A_1$ when
running on $C$  in the sens given by $x_0x_1$. If $i \not = 1$ or $j
\not = 2k-1$ $P=\{x_{i-1},x_i\}$ and $P'=\{x_j,x_{j+1}\}$ are two
distinct balanced pairs intersecting $A_1$ exactly once each. Assume
that $i=1$ and $j=2k-1$. Since $A_2$ contains another vertex $x_l$
($1 < l <2k-1$). Let $x_m$ be the first vertex in $A_1$ when running
from $x_l$ to $x_{2k-1}$ ($l <m \leq 2k-1$. Then $P=\{x_0,x_1\}$ and
$P'=\{x_{m-1},x_m\}$ are two disjoint balanced pairs intersecting
$A_1$ exactly once each.
\end{Prf}

\begin{Lem}\label{Lemma:DisjointBalancedPairsSpecialEvenCycle}  Let
$C$ be an even cycle of length $2p \geq 8$ and let $x$ and $y$ be
two vertices. Assume that the vertices of $C-\{x,y\}$ are
partitioned into $A$ and $B$ with $|A| \geq p-2$ and $|B| \geq p-2$.
Then there are at least two disjoint balanced pairs intersecting $A$
and $B$ exactly once each.
\end{Lem}
\begin{Prf}
Let us colour alternately the vertices of $C$ in red and blue. If
$A$ contains at least two red (or blue) vertices $u$ and $v$ and $B$
two blue (or red respectively) vertices $u'$ and $v'$ then
$P=\{u,u'\}$ and $P'=\{v,v'\}$ are two disjoint balanced pairs. If
$A$ contains a red vertex $u$ and a blue vertex $v$ and,
symmetrically, $B$ contains a red vertex $u'$ and a blue vertex $v'$
then $P=\{u,u'\}$ and $P'=\{v,v'\}$ are two disjoint balanced. It is
clear that at least one of the above cases must happens and the
result follows.
\end{Prf}

\section{Applications}

From now on, we consider that our graphs are cubic, connected and
bridgeless (multi-edges are allowed). Moreover we suppose that they
are not $3-$edge colourable. Hence these graphs have perfect
matchings and any 2-factor have a non null even number of odd
cycles. If $X \subset V(G)$ and $Y \subset V(G)$, $d(X,Y)$ is the
length of a shortest path between these two sets.

\subsection{Graphs with small oddness}
\begin{Thm} \label{Theorem:Distance3Oddness2}Let $G$ be a cubic graph of oddness two. Assume that $G$
has a perfect matching $M$ where the $2-$factor $\mathcal
C=\{C_1,C_2 \ldots C_k\}$ of $G-M$ is such that $C_1$ and $C_2$ are
the only odd cycles and $d(C_1,C_2) \leq 3$. Then $G$ has three
perfect matchings with an empty intersection.
\end{Thm}
\begin{Prf}

If $d(C_1,C_2)=1$ let $uv$ be an edge joining $C_1$ and $C_2$ ($u
\in C_1$ and $v \in C_2$). $A=\{uv\}$ is a balanced $M-$matching.
Let $M_2$ be a perfect matching such that $M_2 \cap M= A$. There is
certainly a perfect matching $M_3$ avoiding $uv$ (see Theorem
\ref{Theorem:EdmondsSeymour}). Hence $M$, $M_1$ and $M_3$ are three
perfect matchings with an empty intersection.

It can be noticed that $d(C_1,C_2) \not = 2$. Indeed, Let
$P=u_1vu_2$ be a shortest path joining $u_1 \in C_1$ to $u_2 \in
C_2$, then the cycle of $\mathcal C$ containing $v$ cannot be
disjoint from $C_1$ or $C_2$, impossible.

Assume thus now that $d(C_1,C_2)  = 3$ and let $P=u_1u_2u_3u_4$ be a
shortest path joining $C_1$ to $C_2$ (with $u_1 \in C_1$ and $u_4
\in C_2$). Then $A=\{u_1u_2,u_3u_4\}$ is a balanced $M-$matching.
Let $M_2$ be a perfect matching such that $M_2 \cap M= A$. From
Lemma \ref{Lemma:TwoEdgesAvoiding} there is a perfect matching $M_3$
avoiding these two edges of $A$. Hence $M$, $M_2$ and $M_3$ are
three non intersecting perfect matchings
\end{Prf}

A graph $G$ is {\em near-bipartite} whenever there is an edge $e$ of
$G$ such that $G-e$ is bipartite.

\begin{Thm} \label{Theorem:3cyclesOddness2}Let $G$ be a cubic graph of oddness two. Assume that $G$
has a perfect matching $M$ where the $2-$factor $\mathcal C=$ of
$G-M$ has only $3$ cycles $C_1,C_2$ (odds) and $ C_3$ (even) such
that the subgraph of $G$ induced by $C_3$ is a near-bipartite graph.
Then $G$ has three perfect matchings with an empty intersection.
\end{Thm}
\begin{Prf}
From Theorem \ref{Theorem:Distance3Oddness2}, we can suppose that
$d(C_1,C_2) \geq 3$. That means that the neighbors of $C_1$ are
contained in $C_3$ as well as those of $C_2$. Let us colour the
vertices of $C_3$ with two colours red and blue alternately along
$C_3$. Assume that $a$ and $b$ are two vertices of $C_3$ with
distinct colours such that $a$ is a neighbor of $C_1$ and $b$ is a
neighbor of $C_2$. Let $e$ and $f$ be the two edges of $M$ so
determined by $a$ and $b$. Then $A=\{e,f\}$ is a balanced
$M-$matching. Let $M_2$ be a perfect matching such that $M \cap
M_2=A$ and $M_3$ be a perfect matching avoiding $A$ (Lemma
\ref{Lemma:TwoEdgesAvoiding}). Then $M,M_1$ and $M_2$ are $3$ non
intersecting perfect matchings.

It remains thus to assume that the neighbors of $C_1$ and $C_2$ have
the same colour (say red). $G$ being bridgeless, we have an odd
number (at least $3$) of edges in $M$ joining $C_1$ and $C_3$ ($C_2$
and $C_3$ respectively). The remaining vertices of $C_3$ are matched
by edges of $M$, but we have at least $6$ blue vertices more than
red vertices in $C_3$ to be matched and hence at least three pairs
of blue vertices must be matched. Let $e \in E(G)$ such that $G-e$
is bipartite, if $e\in C_3$ then $C_3$ must have odd length,
impossible. Hence $e$ is the only chord of $C_3$ whose ends have the
same colour, impossible.

\end{Prf}
\begin{Thm} \label{Theorem:Oddness4Distance1}
Assume that $G$ is a cubic graph having a perfect matching $M$ where
the $2-$factor $\mathcal C=\{C_1,C_2,C_3,C_4 \ldots C_k\}$ of $G-M$
is such that $C_1$, $C_2$, $C_3$ and $C_4$ are the only odd cycles.
Assume moreover that $d(C_1,C_2) =1$ as well as $d(C_3,C_4)=1$. Then
$G$ has three perfect matchings with an empty intersection.
\end{Thm}
\begin{Prf}
Let $u_1u_2$ be an edge joining $C_1$ to $C_2$ and $u_3u_4$ be an
edge joining $C_3$ to $C_4$. $A=\{u_1u_2,u_3u_4\}$ is a balanced
$M-$matching. Let $M_2$ be a perfect matching such that $M \cap M_2
= A'$. By Lemma \ref{Lemma:TwoEdgesAvoiding}, there is a perfect
matching $M_3$ avoiding these two edges. Hence the three perfect
matchings $M$, $M_2$ and $M_3$ are non intersecting.
\end{Prf}

\begin{Thm} \label{Theorem:4OddChordlessCycles}
Assume that $G$ has a perfect matching $M$ where the $2-$factor
$\mathcal C$ has only $4$ chordless cycles $\mathcal
C=\{C_1,C_2,C_3,C_4\}$. Then $G$ has three perfect matchings with an
empty intersection.
\end{Thm}
\begin{Prf}
By the connectivity of $G$, every vertex of three cycles of
$\mathcal C$ (say $C_1,C_2$ and $C_3$) are joined to $C_4$ while no
other edge exists. Otherwise the result holds by Theorem
\ref{Theorem:Oddness4Distance1}.

Each cycle of $\mathcal C$ has length at least $3$ and, hence $C_4$
has length at least $9$. We can colour each vertex $v \in C_4$ with
$1$, $2$ or $3$ following the fact the edge of $M$ incident with $v$
has its other end on $C_1$, $C_2$ or $C_3$. From lemma
\ref{Lemma:DisjointBalancedTriples}, there is two balanced triples
$T$ and $T'$ intersecting  each colour. These two balanced triples
determine two disjoint balanced $M-$matchings. Hence, the result
holds from Lemma \ref{Lemma:FondamentalDisjointsBalanced}.
\end{Prf}

\subsection{Good Rings, Good stars}

A {\em good path of index $C_0$} is a set $P$ of $k+1$ disjoint
cycles $C_0,C_1 \ldots C_k$ such that

\begin{itemize}
  \item $C_0$ and $C_k$ are the only odd cycles of $P$
  \item $C_i$ is joined to $C_{i+1}$ ($0 \leq i \leq k-1$)  by an
  edge $e_i$ (called {\em jonction edge of index $C_0$})
  \item the two jonction edges incident to an even cycle
  determine two odd paths on this cycle
\end{itemize}

A {\em good ring} is a set $R$ of disjoint odd cycles $C_0 \ldots
C_{2p-1}$ and even cycles such that
\begin{itemize}
  \item $C_i$ is joined to $C_{i+1}$ ($i$ is taken modulo $2p$) by a
  good path $P_i$ of index $C_i$ whose even cycles are in $R$
  \item the good paths involved in $R$ are pairwise disjoint.
\end{itemize}

A {\em good star} ({\em centered in $C_0$}) is a set $S$ of four
disjoint cycles $C_0,C_1,C_2,C_3$ such that
\begin{itemize}
  \item $C_0$ (the center) is chordless and has length at least $7$
  \item $C_0$ is joined to each other cycle by at
  least two edges and has no neighbor outside of $S$
  \item there is no edge between $C_1$, $C_2$ and $C_3$
\end{itemize}

\begin{Thm} \label{Theorem:GoodRingGoodStars}
Assume that $G$ has a perfect matching $M$ where the $2-$factor
$\mathcal C$ of $G-M$ can be partitioned into good rings, good stars
and even cycles. Then $G$ has three perfect matchings with an empty
intersection.
\end{Thm}
\begin{Prf}
Let $\mathcal R$ be the set of good rings of $\mathcal C$ and
$\mathcal S$ be the set of good stars.

Let $R \in \mathcal R$, and let $C_0 \ldots C_{2p-1}$ be its set of
odd cycles.  Let us us say that a junction edge of $R$ has an even
index whenever this edge is a junction edge of index $C_i$ with $i$
even. A junction edge of odd index is defined in the same way. Let
$A_R$ be the set of junction edge of even index of $R$ and $B_R$ the
set of junction edge of odd index. We let $A=\bigcup_{R \in \mathcal
R} A_R$ and $B=\bigcup_{R \in \mathcal R} B_R$.

For each star $S \in \mathcal S$, assume that each vertex of the
center is coloured with the name of the odd cycle of $S$ to whom
this vertex is adjacent.  Let $T_S$ and $T'_S$ be two disjoint
balanced triples (Lemma \ref{Lemma:DisjointBalancedTriples}) of the
center of $S$. Let $N_S$ and $N'_S$ be the sets of three edges
joining the center of $S$ to the other cycles of $S$, determined by
$T_S$ and $T'_S$. Let $A' = \bigcup_{S \in \mathcal S} N_S$ and
$B'=\bigcup_{S \in \mathcal S}N'_S$.

It is an easy task to check that $A + A'$ and $B+B'$ are two
disjoint balanced $M-$matchings. Hence, the result holds from Lemma
\ref{Lemma:FondamentalDisjointsBalanced}.
\end{Prf}

A particular case of the above result is given by E.
M\`{a}\v{c}ajov\'{a} and M. \v{S}koviera. The length of a ring is
the number of jonction edges. A ring of length $2$ is merely a set
of two odd cycles joined by two edges.
\begin{Cor}\cite{MacSko}Assume that $G$ has a perfect matching $M$ where the odd cycles of the $2-$factor
$\mathcal C$ can be arranged into rings of length $2$. Then $G$ has
three perfect matchings with an empty intersection.
\end{Cor}

It can be pointed out that this technique of rings of length $2$ was
used in \cite{Fou85} for the $5-$ flow problem when dealing with
graphs of small order and graphs with low genus. This technique has
been developped independently  by Steffen in \cite{Ste96}.

\section{On graphs with at most $32$ vertices}

Determining the structure of a minimal counterexample to a
conjecture is one of the most typical methods in Graph Theory. In
this section we investigate some basic structures of minimal
counterexamples to Conjecture \ref{Conjecture:FanRaspaud}.

The {\em girth} of a graph is the length of shortest cycle.
M\`{a}\v{c}ajov\'{a} and  \v{S}koviera \cite{MacSko} proved that the
girth of a minimal counterexample is at least $5$.
\begin{Lem}\cite{MacSko} \label{Lemma:Girth5}  If $G$ is a smallest bridgeless
cubic graph with no $3$ non-intersecting perfect matchings, then the
girth of $G$ is at least $5$
\end{Lem}

\begin{figure}[t]
\begin{center}
\centering\epsfsize=0.30 \hsize\epsfbox {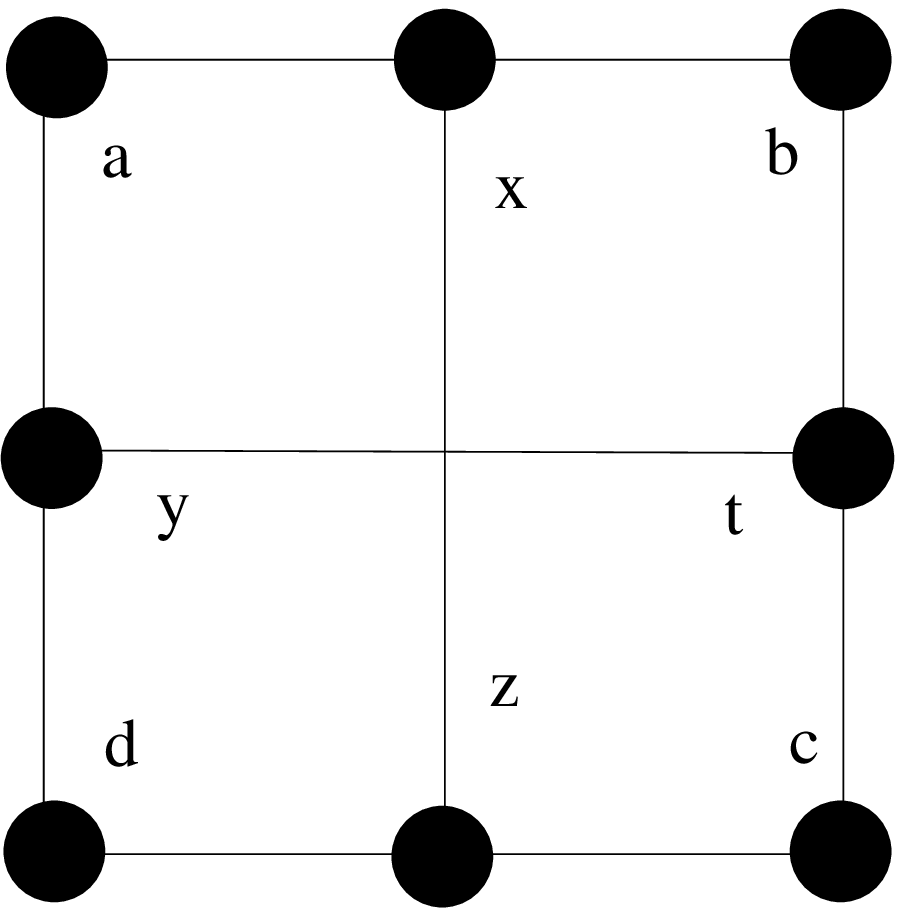} \caption{$G_8$}
 \label{Figure:G8}
\end{center}
\end{figure}

\begin{Lem}\label{Lemma:Graph8}  If $G$ is a smallest bridgeless
cubic graph with no $3$ non-intersecting perfect matchings, then $G$
does not contain a subgraph isomorphic to $G_8$  (see Figure
\ref{Figure:G8}).
\end{Lem}
\begin{Prf}
Assume that $G$ contains $G_8$. Let $a', b', c'$ and $d'$ be the
vertices of $G-G_8$ adjacent to, respectively $a, b, c$ and $d$. Let
$G'$ be the graph obtained in deleting $G_8$ and joining $a'$ to
$c'$ and $b'$ to $d'$. It is an easy task to verify that $G'$ has
chromatic index $3$ if and only if $G$ itself has chromatic index
$3$.   We do not know whether this graph is connected or not but
each component is smaller than $G$ and contains thus $3$
non-intersecting perfect matchings leading to $3$ non-intersecting
perfect matchings for $G'$. Let $P_1$ $P_2$ and $P_3$ these perfect
matchings. Our goal is to construct $3$ non-intersecting perfect
matchings for $G$ $M_1$, $M_2$ and $M_3$ from those of $G'$. We have
thus to delete the edge $a'c'$ and $b'd'$ from $P_1$, $P_2$ and
$P_3$ whenever they belong to these sets and add some edges of $G_8$
in order to obtain the perfect matchings for $G$.

%
Let us now consider the number of edges in $\{a'c',b'd'\}$ which are
contained in $P_1\cap P_2$ or in $P_1\cap P_3$ or in $P_2\cap P_3$.

When none of $P_1\cap P_2$,$P_1\cap P_3$ or $P_2\cap P_3$ contain
$a'c'$ nor $b'd'$ we set $M_1=P_1+\{ax, bt, cz, dy\}$, $M_2=P_2+\{
ay, dz, ct, bx\}$ and $M_3=P_3+\{ax, bt, cz, dy\}$.

Assume that the edges $a'c'$ and $b'd'$ both belong to some $P_i\cap
P_j$ ($i\neq j\in\{1,2,3\}$), say $P_1\cap P_2$. In this case $P_3$
cannot contain one of those edges. Thus we write
$M_1=P_1-\{a'c',b'd'\}+\{a'a,c'c,b'b,d'd\}+\{xz,
ut\}$,$M_2=P_2-\{a'c',b'd'\}+\{a'a,c'c,b'b,d'd\}+\{xz, ut\}$ and
$M_3=P_3+\{ax, bt, cz, dy\}$.

Finally assume w.l.o.g that $P_1\cap P_2=\{a'c'\}$. When $P_2\cap
P_3=P_1\cap P_3=\empty set$ we set $M_1=P_1-\{a'c'\}+\{a'a,
c'c\}+\{yt, xb, dz\}$, $M_2=P_2-\{a'c'\}+\{a'a, c'c\}+\{bt, xz,
dy\}$ and $M_3=P_3+\{ax, bt, cz, dy\}$. On the last hand, if one of
the sets $P_2\cap P_3$ or $P_1\cap P_3$ (say $P_2\cap P_3$) contain
the edge $b'd'$, we write $M_1=P_1-\{a'c'\}+\{a'a, c'c\}+\{yt, xb,
dz\}$,$M_2=P_2-\{a'c',b'd'\}+\{a'a, b'b,c'c, d'd\}+\{xz, yt\}$ and
$M_3=P_3-\{b'd'\}+\{b'b,d'd\}+\{ay,xz,ct\}$.

In all cases, since $P_1\cap P_2\cap P_3=\emptyset$ we have $M_1\cap
M_2\cap M_3=\emptyset$.

\end{Prf}

\begin{Lem}\label{Lemma:P-vExcluded}  If $G$ is a smallest bridgeless
cubic graph with no $3$ non-intersecting perfect matchings, then $G$
does not contain a subgraph isomorphic to the Petersen graph with
one vertex deleted.
\end{Lem}
\begin{Prf}
Let $P$ be a graph isomorphic to the Petersen graph whose vertex set
is $\{a,b,c,d,e,x,y,z,t,u\}$ and such that $abcde$ and $xyztu$ are
the two odd cycles of the $2$-factor associated to the perfect
matching $\{ax, bt, cy, du, ez\}$. Assume that $H=P-a$ is a subgraph
of $G$. Let $x'$, $b'$ and $c'$ be respectively the neighbors of
$x$, $b$ and $c$ in $G-H$. Let $G'$ be the graph whose vertex set is
$V(G-H)\cup\{v\}$ where $v\notin V(G)$ is a new vertex and whose
edge set is $E(G-H)\cup\{vx', ve', vb'\}$. Since $G'$ is smaller
than $G$, $G'$ contains $3$ non-intersecting perfect matchings
$P_1$, $P_2$, $P_3$.

For $i\in\{1,2,3\}$ we can associate to $P_i$ two perfect matchings
of $G$ , namely $M_i$ and $M'_i$, as follows (observe that exactly
one of the edges $vx'$, $vb'$ or $vc'$ belongs to $P_i$)~:
\begin{description}
\item When $vx'\in P_i$ we set $M_i=P_i-\{vx'\}\cup\{xx', bt, cy, du, ez\}$ \\and $M'_i=P_i-\{vx'\}\cup\{xx', tu, bc, yz, ed\}$.
\item When $vb'\in P_i$ we set $M_i=P_i-\{vb'\}\cup\{bb', cy, xu, de, zt\}$ \\ and $M'_i=P_i-\{vb'\}\cup\{bb', cd, ut, ez, xy\}$.
\item When $ve'\in P_i$ we set $M_i=P_i-\{ve'\}\cup\{ee', cd, bt, zy, xu\}$ \\ and $M'_i=P_i-\{ve'\}\cup\{ee', du, xy, zt, bc\}$.
\end{description}
But now, if on one hand $P_i\cap P_j$ contains one of the edges in
$\{vx', vb', ve'\}$  for some $i\neq j\in\{1,2,3\}$ and  for $k\in
\{1,2,3\}$ distinct from $i$ and $j$, $M_i\cap M'_j\cap M_k=M_i\cap
M'_j\cap M'_k=P_1\cap P_2\cap P_3=\emptyset$,a contradiction. If, on
the other hand, each of $P_i$, $P_j$ and $P_k$ (for $i$, $j$, $k$
distinct members of $\{1,2,3\}$) contains exactly one edge of
$\{vx',vb',vc'\}$ we also have $M_i\cap M_j\cap M_k=P_i\cap P_j\cap
P_k=\emptyset$, a contradiction.
\end{Prf}
\begin{Thm} \label{Theorem:MinimumCounterExample32}
If $G$ is a smallest bridgless cubic graph with no $3$
non-intersecting perfect matchings, then $G$ has at least $32$
vertices
\end{Thm}
\begin{Prf}
Assume to the contrary that $G$ is a counterexample with at most
$30$ vertices. We can obviously suppose that $G$ is connected. Let
$M$ be a perfect matching and let $\mathcal C$ be the $2-$factor of
$G-M$. Assume that the number of odd cycles of $\mathcal C$ is the
oddness of $G$.  Since $G$ has girth at least $5$ by Lemma
\ref{Lemma:Girth5}, the oddness of $G$ is $2$, $4$ or $6$.
\begin{Clm} \label{Claim:Claim1MinimummCounterExample32}
$G$ has oddness $2$ or $4$.
\end{Clm}
\begin{PrfClaim} Assume  that $G$ has oddness $6$. We have
$\mathcal C=\{C_1,C_2,C_3,C_4,C_5,C_6\}$ and each cycle $C_i$ ($i=1
\ldots 6$) is chordless and has length $5$. Each cycle $C_i$ is
joined to at least two other cycles of $\mathcal C$. Otherwise, if
$C_i$ is joined to only one cycle $C_j$ ($i \not = j$), these two
cycles would form a connected component of $G$ and  $G$ would not be
connected, impossible. It is an easy task to see that we can thus
partition $\mathcal C$  into good rings and the results comes from
Theorem \ref{Theorem:GoodRingGoodStars}.
\end{PrfClaim}

Assume now that $G$ has oddness $4$. Hence $\mathcal C$ contains $4$
odd cycles $C_1,C_2,C_3$ and $ C_4$. Since these cycles have length
at least $5$, $\mathcal C$ contains eventually an even cycle $C_5$.
From Lemmas \ref{Lemma:Girth5} and  \ref{Lemma:Graph8}
 if $C_5$ exists,  $C_5$ is a chordless cycle of length $6$ or $C_5$ has length
$8$ (with at most one chord) or $10$. When $C_5$ has length $10$,
$C_1,C_2,C_3$ and $C_4$ are chordless cycles of length $5$. When
$C_5$ has length $8$, $C_1,C_2,C_3$ and $C_4$ are chordless cycles
of length $5$ or $3$ of them have length $5$ while the last one has
length $7$.

Theorem \ref{Theorem:Oddness4Distance1} says that we are done as
soon as we can  find two edges allowing to arrange by pairs
$C_1,C_2,C_3$ and $C_4$  (say for example $C_1$ joined to $C_2$ and
$C_3$ to $C_4$)  and Theorem \ref{Theorem:GoodRingGoodStars} says
that we are done whenever these $4$ odd cycles induce a good star.
That means that the subgraph $H$ induced  by the four odd cycles is
of one of the two following types:

\begin{itemize}
  \item [Type 1]One odd cycle (say $C_4$) has all its neighbors in $C_5$ and the $3$
other odd cycles induce a connected subgraph
  \item [Type 2] One cycle (say
$C_4$) is joined to the other by at least one edge while the others
are not adjacent.
\end{itemize}

\begin{Clm} \label{Claim:Claim2_1_MinimumCounterExample32}
$C_5$ has length at least $8$.
\end{Clm}
\begin{PrfClaim}
Assume that $|C_5|=6$, the girth of $G$ being at least $5$ (Lemma
\ref{Lemma:Girth5}) we can suppose that $C_5$ has no chord. $H$ is
not of type $1$, otherwise $C_4$ having its neighbors in $C_5$,
$C_5$ is connected to the remaining part of $G$ with one edge only,
impossible since $G$ is bridgeless. Assume thus that $H$ is of type
$2$. Then, there are $6$ edges between $C_5$ and $H$. Since there
are at least $15$ edges going out $C_1,C_2$ and $C_3$ that means
that there are at least $9$ edges between $C_0$ and the other odd
cycles. Hence, $C_0$ must have length $9$ and can not be adjacent to
$C_5$. $G$ is then partitioned into a good star and an even cycle
and the result comes from Theorem \ref{Theorem:GoodRingGoodStars}.
\end{PrfClaim}

\begin{Clm} \label{Claim:Claim2_2_MinimumCounterExample32}
If $C_5$ has length $8$ then it has no chord.
\end{Clm}
\begin{PrfClaim} If $C_5$ has a chord then there are at most $6$
edges joining $C_5$ to $H$. If $H$ is of type $1$ then $C_4$ has at
least $5$ neighbors in $C_5$. Hence there is at most one edge
between $H$ and $C_5$, impossible. If $H$ is of type $2$, then the
the three cycles $C_1$, $C_2$ and $C_3$ have at least $9$ neighbors
in $C_4$, impossible since $G$ has at most $30$ vertices.
\end{PrfClaim}

\begin{Clm} \label{Claim:Claim3MinimumCounterExample32}
If $C_5$ exists then $H$ is not of type 1.
\end{Clm}
\begin{PrfClaim}
If $H$ is of type 1, then $C_4$ has its neighbors (at least $5$) in
$C_5$ and there are $3$ or $5$ edges between $H$ and $C_5$.

Whenever there are $5$ edges between $H$ and $C_5$, $C_5$ has length
$10$ and $C_1,C_2,C_3$ have length $5$ (as well as $C_4$). In that
case w.l.o.g., we can consider that $C_3$ is joined by exactly one
edge to $C_5$ and joined by $4$ edges to $C_2$. The las t neighbor
of $C_2$ cannote be on $C_5$, otherwise the $5$ neighbors of $C_1$
are on $C_5$ and $C_5$ must have length $12$, impossible. Hence,
$C_2$ is joined to $C_1$ by exactly one edge and $C_1$ is joined to
$C_5$ by $4$ edges. Let us colour each vertex $v$ of $C_5$ with
$1,3$ or $4$ when $v$ is adjacent to $C_i$ ($i=1,3,4$). From Lemma
\ref{Lemma:DisjointBalancedPairs}, we can find $2$ disjoint balanced
pairs on $C_5$ $P=\{u,v\}$ and $P'=\{u',v'\}$ with $u$ and $u'$
coloured with $4$, $v$ and $v'$ coloured with $1$. These two pairs
determine two disjoint set of edges $N'=\{e,f\}$ and $N"=\{h,i\}$ in
$M$ and allow us to construct two disjoint balanced $M-$matchings
$M'=\{e,f,g\}$ and $M"=\{h,i,j\}$ in choosing two distinct edges $g$
and $j$ between $C_2$ and $C_3$. The result follows from Lemma
\ref{Lemma:FondamentalDisjointsBalanced}

Whenever there are $3$ edges between $H$ and $C_5$, $C_5$ has length
$8$ or $10$, any two cycles in $\{C_1,C_2,C_3\}$ are joined by at
least two edges and  each of them is joined to $C_5$ by exactly one
edge. Let $A$ be the three vertices of $C_5$ which are the neighbors
of $C_1 \cup C_2 \cup C_3$. Let $B$ be the neighbors of $C_4$ on
$C_5$. When $C_5$ has length $10$ this cycle induces a  chord $xy$.
In that case, Lemma
\ref{Lemma:DisjointBalancedPairsSpecialEvenCycle} says that we can
find $2$ disjoint balanced pairs $P=\{u,v\}$ and $P'=\{u',v'\}$ with
$u,u' \in A$  and  $v,v' \in B$. These two pairs determine two
disjoint set of edges $N'=\{e,f\}$ and $N"=\{h,i\}$ in $M$ and allow
us to construct two disjoint balanced $M-$matchings $M'=\{e,f,g\}$
and $M"=\{h,i,j\}$ in choosing two suitable distinct edges $g$ and
$j$ joining two of the cycles in $\{C_1,C_2,C_3\}$. When $C_5$ has
no chord, we can apply the same technique in choosing $x$ and $y$ in
$B$.

The result follows from Lemma
\ref{Lemma:FondamentalDisjointsBalanced}.

\end{PrfClaim}

\begin{Clm} \label{Claim:Claim4_1_MinimumCounterExample32}
if  $H$ is  of type 2 then $C_5$ has $8$ vertices.
\end{Clm}
\begin{PrfClaim} When $C_5$ has length $10$, this cycle has no chord. Otherwise, we
have at most $8$ edges between $H$ and $C_5$. Hence $C_1,C_2$ and
$C_3$ are joined to $C_4$ with at least $7$ edges, impossible since
$G$ hat at most $30$ vertices. Assume thus that $C_5$ is a chordless
cycle of length $10$ then there are $15$ edges going out $C_1 \cup
C_2 \cup C_3$ and at most $5$ of them are incident to $C_4$. Hence
there are $10$ edges between $C_1 \cup C_2 \cup C_3$ and $C_5$, $5$
edges between $C_1 \cup C_2 \cup C_3$ and $C_4$ and henceforth no
edge between $C_4$ and $C_5$. One cycle in $\{C_1,C_2,C_3\}$ has
exactly one neighbor in $C_5$ (say $C_1$) or two of them (say $C_1$
and $C_2$) have this property .

It is an easy task to find a balanced triple $u,v,w$ on $C_4$ where
$u$ is a neighbor of $C_1$, $v$ a neighbor of $C_2$ and $w$ a
neighbor of $C_3$. This balanced triple determine a balanced
$M-$matching $A$. We can construct a balanced $M-$matching $B$
disjoint from $A$ in choosing two edges $e$ end $f$ connecting $C_1
\cup C_2$ to $C_5$ whose ends are adjacent on $C_5$ (since $7$ or
$8$ edges are involved between these two sets) and an edge $h \not
\in A$ between $C_3$ and $C_4$. The result follows from Lemma
\ref{Lemma:FondamentalDisjointsBalanced}

\end{PrfClaim}

\begin{Clm} \label{Claim:Claim4_2_MinimumCounterExample32}
If $C_5$ exists then $H$ is  not of type 2.
\end{Clm}
\begin{PrfClaim} From claim \ref{Claim:Claim4_1_MinimumCounterExample32}, it remains to assume that
 $C_5$ has length $8$. Then $C_1 \cup C_2 \cup C_3$ is joined to
$C_4$ by at least $7$ edges. $C_4$ has then no neighbor in $C_5$ and
$G$ is partitioned into a good star centered on $C_4$ and an even
cycle as soon as $C_1$, $C_2$ and $C_3$ have two neighbors at least
in $C_4$. In that case, the result follows from Theorem
\ref{Theorem:GoodRingGoodStars}.

Assume thus that $C_1$ has only one neighbor in $C_4$ (and then $4$
neighbors in $C_5$). Assume that $C_2$ has more neighbors in $C_5$
than $C_3$. Hence $C_2$ has at least $2$ neighbors in $C_5$. Let us
colour each vertex $v$ of $C_5$ with $1,2$ or $3$ when $v$ is
adjacent to $C_1,C_2$ or $C_3$. With that colouring $C_5$ is a good
even cycle. We can find $2$ disjoint balanced pairs intersecting the
colour $1$ exactly once each. Let $\{e,f\}$ and $\{g,h\}$ the two
pairs of edges of $M$ so determined. We can complete these two pairs
with a third edge $i$ ($j$ respectively) connecting $C_3$ to $C_4$
or $C_2$ to $C_3$,following the cases, in such a way that
$A=\{e,g,i\}$ and $B=\{f,h,j\}$ are two disjoint balanced
$M-$matchings. The result follows from Lemma
\ref{Lemma:FondamentalDisjointsBalanced}

\end{PrfClaim}

\begin{Clm} \label{Claim:Claim5MinimumCounterExample32}
The oddness of $G$ is at most $2$.
\end{Clm}
\begin{PrfClaim} In view of the previous claims, it remains to consider the case were $\mathcal C$ is reduced to a
set of four odd cycles $\{C_1,C_2,C_3,C_4\}$. Once again, Theorem
\ref{Theorem:Oddness4Distance1}, says that, up to the name of
cycles, $C_1,C_2$ and $C_3$ are joined to the last cycle $C_4$ and
have no other neighboring cycle. That means that $C_1$, $C_2$, $C_3$
have length $5$ and $C_4$ has length $15$. These $4$ cycles are
chordless and the result comes from Theorem
\ref{Theorem:4OddChordlessCycles}.
\end{PrfClaim}

Hence, we can assume that $\mathcal C$ contains only two odd cycles
$C_1$ and $C_2$. Since we consider graphs with at most $30$ vertices
and since the even cycles of $\mathcal C$ have length at least $6$,
$\mathcal C$ contains only one even cycle $C_3$ or two even cycles
$C_3$ and $C_4$ or three even cycles $C_3,C_4$ and $C_5$. From
Theorem \ref{Theorem:Distance3Oddness2}, $C_1$ and $C_2$ are at
distance at least $4$. That means that the only neighbors of these
two cycles are vertices of the remaining even cycles.

It will be convenient, in the sequel, to consider  that the vertices
of the even cycles are coloured alternately in red and blue.

\begin{Clm} \label{Claim:Claim6MinimumCounterExample32}
If $C_1$ and $C_2$ are  joined to an even cycle in $\mathcal C$,
then their neighbors in that even cycle have the same colour
\end{Clm}
\begin{PrfClaim}
 Assume
that $C_1$ is joined to a blue vertex of an even cycle of $\mathcal
C$ by the edge $e$ and $C_2$ is joined to a red vertex of this same
cycle  by the edge $e'$. $A=\{e,e'\}$ is then a balanced
$M-$matching. Let $M_2$ be the perfect matching of $G$ such that $M
\cap M_2= A$  and let $M_3$ be a perfect matching avoiding $e$ and
$e'$ (Lemma \ref{Lemma:TwoEdgesAvoiding}). then $M$, $M_2$ and $M_3$
are two non intersecting perfect matchings, a contradiction.
\end{PrfClaim}

Hence, for any even cycle of $\mathcal C$  joined to the two cycles
$C_1$ and $C_2$, we can consider that, after a possible permutation
of colours for some even cycle,  the vertices adjacent to $C_1$ or
$C_2$ have the same colour (say red).

\begin{Clm} \label{Claim:Claim7MinimumCounterExample32}
$\mathcal C$ contains $2$ even cycles
\end{Clm}
\begin{PrfClaim}
Assume that $\mathcal C$ contains $3$ even cycles $C_3,C_4$ and
$C_5$. We certainly have, up to isomorphism, $C_3$ and $C_4$ with
length $6$ and $C_5$ of length $6$ or $8$ while the lengths of $C_1$
and $C_2$ are bounded above by $7$. In view of Claim
\ref{Claim:Claim6MinimumCounterExample32} $C_1 \cup C_2$ has at most
$3$ neighbors in $C_3$ and in $C_4$ and at most $4$ neighbors in
$C_5$. Since $C_1$ and $C_2$ have at least $10$ neighbors, that
means that all the red vertices of $C_3 \cup C_4 \cup C_5$ are
adjacent to some vertex in $C_1$ or $C_2$. It is then easy to see
that two even cycles are joined  by two distinct edges ($i$ and $j$)
whose ends are blue and each of them is connected to both $C_1$ and
$C_2$ (say $e$ and $f$ connecting  $C_1$ and $g$ and $h$ connecting
$C_2$). Then $A=\{e,g,i\}$ and $B=\{f,h,j\}$ are two disjoint
balanced $M-$matchings and the result follows.

 Assume now that
$\mathcal C= \{C_1,C_2,C_3\}$. Since $C_1$ and $C_2$ have at least
$5$ neighbors each in  $C_3$, $C_3$ must have $10$ red vertices.
Hence $C_1$ and $C_2$ have length $5$ and $C_3$ has length $20$. The
$10$ blue vertices of $C_3$ are matched by $5$ edges of $M$. For any
chord of $C_3$, we can find a red vertex in each path determined by
this chord on $C_3$, one being adjacent to $C_1$ and the other to
$C_2$. Let $A$ be the three edges so determined. $A$ is a balanced
$M-$matching. By systematic inspection we can check that it is
always possible to find two disjoint balanced $M-$matchings so
constructed. The result follows from Lemma
\ref{Lemma:FondamentalDisjointsBalanced}
\end{PrfClaim}

We shall say that $G$ is a graph of {\em type 3} when
\begin{itemize}
  \item [Type 3]$\mathcal C$ contains two even cycles $C_3$ and $C_4$, the
neighborhood of $C_1$ is contained in $C_3$, the neighborhood of
$C_2$ is contained in $C_4$, and $C_3$ and $C_4$ are joined by $3$
or $5$ edges.
\end{itemize}

\begin{Clm} \label{Claim:Claim8MinimumCounterExample32}
$C_1$ and $C_2$ have length $5$ or $7$ or one of them has length
$9$. In the latter case $G$ is a graph of type $3$
\end{Clm}
\begin{PrfClaim}
$G$ being connected and  bridgeless, $C_1$ and $C_2$ are joined to
the remaining cycles of $\mathcal C$ by an odd number of edges (at
least $3$).

Assume that $C_1$ has length at least $11$, then there at least $16$
vertices involved in $C_1 \cup C_2$. Hence, $\mathcal C$ contains
exactly one even cycle. From Claim
\ref{Claim:Claim7MinimumCounterExample32} this is  impossible.

Assume that $C_1$ has length  $9$, then if $C_1$ is connected to the
remaining part of $G$ with $3$ edges, that means that $C_1$ has $3$
chords. Since $G$ has girth at least $5$, $C_1$ induces a subgraph
isomorphic to the Petersen graph where a vertex is deleted. This is
impossible in view of Lemma \ref{Lemma:P-vExcluded}.

Hence $C_1$ is connected to the even cycles of $\mathcal C$ with $5$
edges. If $\mathcal C$ has only one cycle $C_3$, then, in view of
claim this cycle must has length at least $20$, impossible. We can
thus assume that $\mathcal C$ contains two cycles $C_3$ and $C_4$.
Since $C_1 \cup C_2$ contains at least $14$ vertices, $C_3$ and
$C_4$ have length $8$. If $C_1$ and $C_2$ have both some neighbors
in $C_3$, there are at most $4$ such vertices in view of Claim
\ref{Claim:Claim6MinimumCounterExample32}. In that case, the
remaining (at least $6$)neighbors are in $C_4$, impossible since
this forces $C_4$ to have length at least $10$.

Hence $C_1$ has all its neighbors in $C_3$ and $C_2$ all its
neighbors in $C_4$. The perfect matching $M$ forces $C_3$ and $C_4$
to be connected with an odd number ($3$ or $5$) of edges and $G$ is
a graph of type 3, as claimed.
\end{PrfClaim}

From now on, we have  $\mathcal C= \{C_1,C_2,C_3,C_4\}$

\begin{Clm} \label{Claim:Claim9MinimumCounterExample32}
$G$ is not a graph of type $3$
\end{Clm}
\begin{PrfClaim}

Let $f=ab$ and $g=cd$ two edges joining $C_3$ to $C_4$ with $a$ and
$c$ in $C_3$. Whatever is the colour of $a$ and $c$ we can choose
two distinct vertices $u$ and $v$ in the neighboring vertices of
$C_1$ on $C_3$ such that $u$ and $a$ have distinct colours as well
as $v$ and $c$. Let $g$ be the edge joining $C_1$ to $u$ and $h$ the
edge joining $C_1$ to $v$. In the same way, we can find two distinct
vertices $w$ and $x$ in the neighboring vertices of $C_2$ on $C_4$
with the same property relatively to $b$ and $d$ leading to the
edges $g'$ and $h'$.

We can check that $M'=\{f,g,h\}$ and $M"=\{f',g',h'\}$ are two
disjoint balanced $M-$matchings. The result follows from Lemma
\ref{Lemma:FondamentalDisjointsBalanced}

\end{PrfClaim}

\begin{Clm} \label{Claim:Claim10MinimumCounterExample32}
One of $C_1$ or $C_2$ has its neighborhood included in $C_3$ or
$C_4$
\end{Clm}
\begin{PrfClaim}
If $C_1$  and $C_2$ have neighbors in $C_3$ and $C_4$ each, then,
from Claim \ref{Claim:Claim6MinimumCounterExample32}, there are at
least $20$ vertices involved in $C_3 \cup C_4$. Hence $C_1$ and
$C_2$ have length $5$ and $C_3 \cup C_4$ contains exactly $20$
vertices. The $10$ red vertices of $C_3 \cup C_4$ are adjacent to
$C_1$ or $C_2$ and the blue vertices are connected together.

Let $f$ be a chord for $C_3$ and $f'$ be a chord for $C_4$ (whenever
these two chord exist). We can find two red vertices in $C_3$
separated by $f$, one being adjacent to $C_1$ by an edge $g$ while
the other is adjacent to $C_2$ by an edge $h$. Let $M'=\{f,g,h\}$ be
the  set of three edges so constructed.In the same way we get
$M"=\{f',g',h'\}$ when considering $C_4$. $M'$ and $M"$ are two
disjoint balanced $M-$matchings. The result follows from Lemma
\ref{Lemma:FondamentalDisjointsBalanced}

Assume thus that $C_1$ has no chord. That means that we can find two
distinct edges $e$ and $f$ connecting $C_3$ to $C_4$. Let $g$ be an
edge connecting $C_1$ to $C_3$, $h$ be an edge connecting $C_2$ to
$C_4$, $i$ an edge connecting $C_1$ to $C_4$ and $j$ an edge
connecting $C_2$ to $C_3$. Then $M'=\{e,g,h\}$ and $M"=\{f,i,j\}$
are two disjoint balanced $M-$matchings. The result follows from
Lemma \ref{Lemma:FondamentalDisjointsBalanced}
\end{PrfClaim}

We can assume now that $C_2$ has its neighbors contained in $C_4$.
Since $G$ is not of type $3$ by Claim
\ref{Claim:Claim9MinimumCounterExample32}, $C_1$ has some neighbor
in $C_4$. $C_4$ must have length $12$ at least from Claim
\ref{Claim:Claim6MinimumCounterExample32}. This forces $C_3$ to have
length $8$, $C_4$
 length $12$ and $C_1$ and $C_2$ lengths $5$. Moreover, there is one
 edge exactly between $C_1$ and $C_4$ and $2$ or $4$ edges between
 $C_3$ and $C_4$. It is then an easy task to find $M'=\{e,f,g\}$ and
 $M"=\{h,i,j\}$ with $e$ and $h$ connecting $C_1$ and $C_3$, $f$ and $i$ connecting
 $C_3$ and $C_4$, $g$ and $j$ connecting $C_4$ and $C_2$ such that
 $M'$ and $M"$ are two disjoint balanced $M-$matchings. The result follows from
Lemma \ref{Lemma:FondamentalDisjointsBalanced}
\end{Prf}

\section{Conclusion}

A Fano colouring of $G$ is any assignment of points of the Fano
plane $\mathcal F_7$ (see, e.g., \cite{MacSko}) to edges of $G$ such
that the three edges incident with each vertex of $G$ are mapped to
three distinct collinear points of $\mathcal F_7$ . The following
conjecture appears in \cite{MacSko}

\begin{Conj}\cite{MacSko} \label{Conjecture:MacajovaSkoviera} Every
bridgeless cubic graph admits a  Fano colouring which uses at most
four lines.
\end{Conj}

In fact, M\`{a}\v{c}ajov\'{a} and  \v{S}koviera  proved in
\cite{MacSko} that conjecture \ref{Conjecture:FanRaspaud} and
Conjecture \ref{Conjecture:MacajovaSkoviera} are equivalent. Hence,
our results can be immediately translated in terms of the
M\`{a}\v{c}ajov\'{a} and \v{S}koviera conjecture.

%
%
%

\bibliographystyle{plain}

\bibliography{Bibliographie}

\end{document}